\documentclass[ %
 reprint,
showpacs,preprintnumbers,
showkeys
 amsmath,amssymb,
 aps,
]{revtex4-1}
 
\usepackage{graphics}% Include figure files
\usepackage{dcolumn}% Align table columns on decimal point
\usepackage{bm}% bold math

\usepackage[usenames, dvipsnames]{color}
\usepackage{epsfig}
\usepackage{amsmath}

\graphicspath{{figs/}}

\usepackage[mathlines]{lineno}% Enable numbering of text and display math
\usepackage{soul}

\def\mtant{{\mbox{Ta}_2\mbox{O}_5}}
\def\tant{$\mtant$}

\def\ttan{$\mbox{TiO}_2$}

\newcommand{\SI}[2]{\ensuremath{#1\,{\rm #2}}}

\def\rtHz{\sqrt{\rm Hz}}
\def\mrtHz{{\rm m}/\rtHz}

\def\frmrtHz{\frac{\rm m}{\rtHz}}

\definecolor{spring}{rgb}{0.7,0.9,0.7}
\definecolor{brick}{rgb}{0.7,0.2,0.1}
\definecolor{redHL}{rgb}{1.0,0.5,0.5}

\begin{document}

\preprint{ADS/???}

\title{Direct Measurement of Coating Thermal Noise in Optical Resonators}

\author{S. Gras}
\author{M. Evans}
 %\altaffiliation{Massachusetts Institute of Technology, 185 Albany St. NW22-295, 02139 MA, USA}%Lines break automatically or can be forced with 
\affiliation{Massachusetts Institute of Technology, 185 Albany St. NW22-295, 02139 MA, USA}
 \email{sgras@ligo.mit.edu}

%\collaboration{LIGO Collaboration}%\noaffiliation

\date{\today}% It is always \today, today,

\begin{abstract}
The best measurements of space and time currently possible (e.g. gravitational wave detectors
 and optical reference cavities) rely on optical resonators,
 and are ultimately limited by thermally induced fluctuations in
 the reflective coatings which form the resonator.
We present measurements of coating thermal noise in the audio band and show that
 for a standard ion beam sputtered coating, the power spectrum of the noise does
 \emph{not} have the expected power-law behavior.
\end{abstract}

%Use showkeys class option if keyword
\maketitle

%\tableofcontents

%%%%%%%%%%%%%%%%%%%%%%%%%%%%%%%%%%%%%%
\section{Introduction}

High-reflectivity mirrors play an important role in precision optical experiments
 such as gravitational-wave detectors~\cite{0264-9381-24-2-008, PhysRevD.78.102003},
 frequency references~\cite{Ludlow:07,nphoton.2012.217},
 and macroscopic quantum measurements~\cite{1367-2630-11-7-073032, Poot2012273}. 
These mirrors depend on multilayer coatings which are deposited with either physical methods 
(sputtering, pulse laser deposition,  molecular beam epitaxy) or chemical methods (vapor deposition).
While the coating is critical to the optical measurement, Brownian motion in coatings can present a 
limiting noise source due to nonzero mechanical dissipation in the deposited layers.

Ion beam sputtering (IBS) for amorphous coatings and molecular beam epitaxy  for crystalline coatings currently
 produce the lowest mechanical loss~\cite{Cole:2016ct}.
Further reductions in coating thermal noise (CTN), while maintaining high optical quality (low absorption and
 scatter, high uniformity), are of great interest for many experiments 
 (e.g., future gravitational-wave detectors ~\cite{CE2017,Miller2015,ETpaper}). 
 
The CTN level of candidate coating materials is most frequently estimated using measurements of
 their mechanical properties: mechanical quality factors, Young's modulus, and Poisson ratio.
The techniques used to measure these parameters include, among others,
 suspended disks~\cite{crooks_blades, harry_blades},
 clamped cantilevers~\cite{pierro_cantilevers},
 and the gentle nodal suspension~\cite{cesarini_nodal}.
The level of CTN is then calculated from the measured parameters,
although uncertainties in their values can produce significant uncertainty in the CTN estimate.
Moreover, this approach may not capture all the phenomena involved in a multilayer coating. 
A direct measurement of the thermal noise of a multilayer coating is thus an important
complement to the above approach.

In reference \cite{Gras-Yam} we introduced a novel technique that directly measures the CTN
of a high-reflectivity mirror. The technique uses a Fabry-Perot cavity in which three transverse electromagnetic 
(TEM), Hermite-Gaussian modes co-resonate: TEM00, TEM02 and TEM20. These modes probe different areas of the 
sample coating, and CTN appears as a fluctuation in the resonant frequency difference of the two higher-order modes
(see Fig.~\ref{fig:cavity}). In this article we present an improved version of this
experiment which can measure CTN with much higher signal-to-noise ratio and provide new information
on the frequency dependence of CTN.

%%%%%%%%%%%%%%%%%%%%%%%%%%%%%%%%%%%%%%
\section{Experimental Setup}\label{exp}

\def\fFSR{f_{\rm FSR}}
\def\fTMS{f_{\rm TMS}}
\def\Pin{P_{\rm in}}
\def\Pcav{P_{\rm cav}}

\begin{figure}[b]
\includegraphics[scale=0.45]{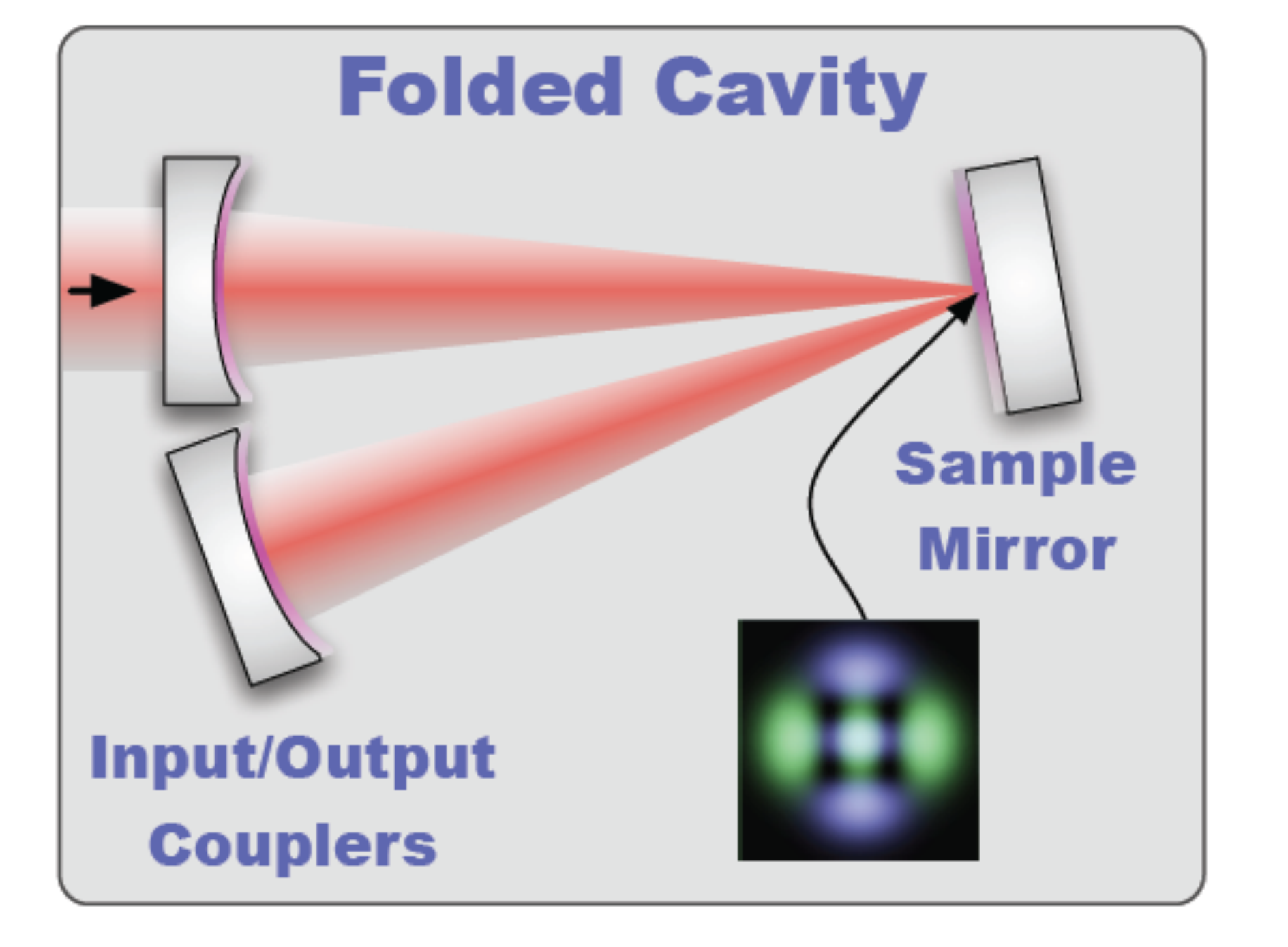}
 \caption{\label{fig:cavity} A high finesse cavity configuration, with a folding
 mirror (the sample to be measured) equidistant from the input and output mirrors.
The inset image shows the TEM20 and TEM02 modes used to make the coating thermal noise measurement.
Since these modes overlap only in a small central area,
 noise in the coating causes changes in the difference between their resonant frequencies,
 while most other noises sources cancel in this difference.} 
\end{figure}

At the core of the experiment is a 3-mirror folded cavity, with the sample to be measured as the folding mirror (see Fig.~\ref{fig:cavity}).
The cavity is mounted on a vibrationally isolated platform in a vacuum chamber ($10^{-5}$\,Torr).
This folded configuration is ideal for rapid testing of high reflectivity coatings, and accepts 
the witness flats commonly included in coating runs.

%%%%%%%%%%
%%%%%%%%%%

%%%%%%%%%%
\begin{figure*}[t!]
\includegraphics[width=0.9\textwidth]{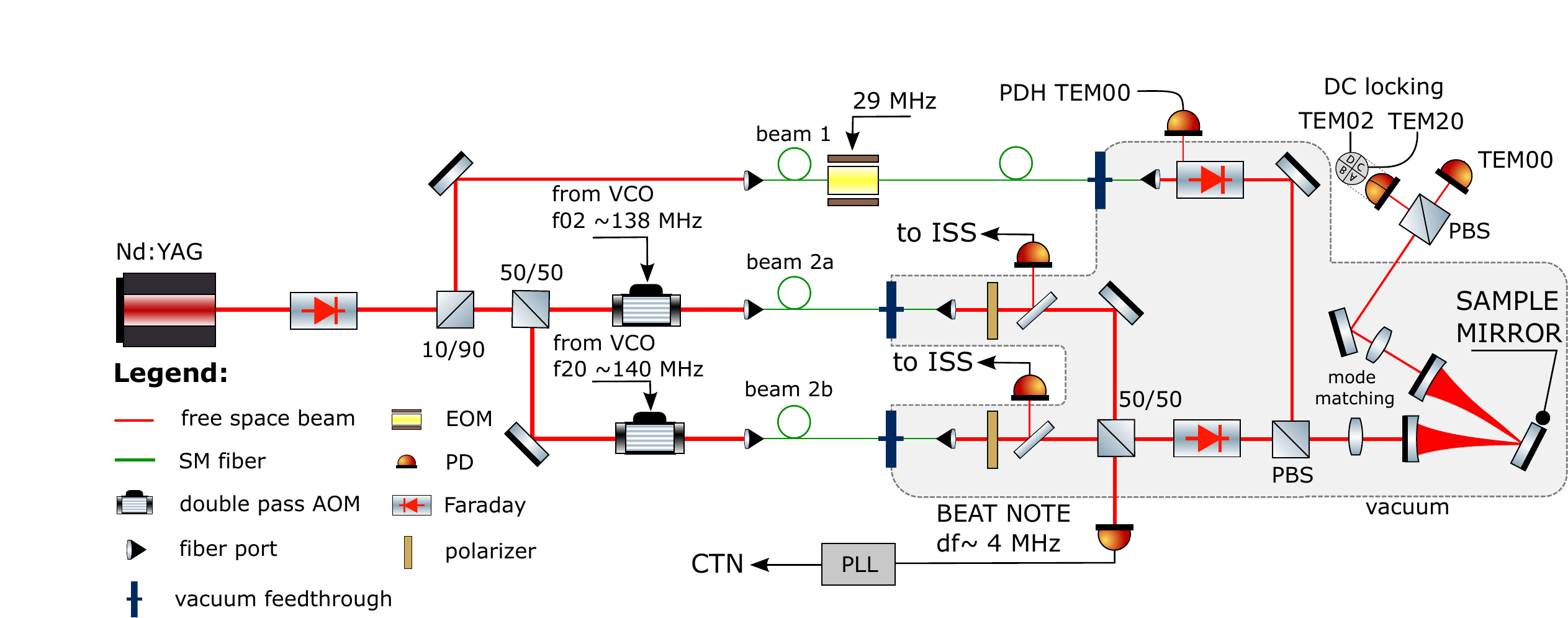}
 \caption{\label{fig:exp} The experimental setup involves a Nd:YAG laser (far left) and an in-vacuum high-finesse cavity (far right). A laser beam is split into 3 paths, 2 of which are shifted in frequency (with AOMs). The laser frequency is controlled to lock the TEM00 mode to the cavity length with PDH locking scheme while the TEM02 and TEM20 modes are DC locked to the cavity. Beams 2a and 2b are intensity stabilized by actuating RF power on AOMs using Intensity Stabilization Servo (ISS) loops.  The primary output of the experiment is the difference between the TEM02 and TEM20 resonant frequencies (labeled BEAT NOTE). Note, beams 1, 2a, 2b are the fundamental TEM00 modes. A conversion of beams 2a, 2b into TEM02 and TEM20 takes place in the cavity.} 

\end{figure*}
%%%%%%%%%%

\begin{table}[b]
\caption{Measured cavity parameters during collection of the data.}
\begin{center}
\begin{tabular}{llcc}
%\hline
Parameter &Symbol & TEM02 & TEM20\\
\hline
Intra-cavity power, W & $\rm{P_{circ}}$ & 2 & 2\\
%{\color{redHL} Transmitted power, mW} & {\color{redHL} $\rm{P_{tr}}$} & {\color{redHL} 0.4} & {\color{redHL}0.4}\\
%{\color{redHL} Cavity pole, kHz} & {\color{redHL} $\nu_{\rm cav}$} & {\color{redHL}50.8} & {\color{redHL} 50.0}\\
 Finesse & $\mathcal{F}$ &15.06 k & 15.30 k	\\
%{\color{redHL} Round trip loss, ppm} &  {\color{redHL} {l}} & {\color{redHL}17} & {\color{redHL} 11}\\
Mode frequency, MHz & $ 2\times \fTMS$ &  276$\pm$2& 280$\pm$2\\
Beam size, $\mu$m  &  $\omega_{S}$ & 54 & 54\\
RoC (effective), mm & R & 50.7 & 50.8\\
%\hline
Laser wavelength, nm & $\lambda$ & \multicolumn{2}{c}{1064} \\
Cavity length, mm & L  &\multicolumn{2}{l}{ $L_1+L_2=46.45+53.07$} \\
Folding angle, deg & $\alpha$ & \multicolumn{2}{c}{17.23}\\
 \hline
\end{tabular}
\end{center}
\label{tab_cav}
\end{table}
%%%%%%%%%%

The cavity is near-concentric, with a total length of $L = \SI{99.5}{mm}$
 and input and output couplers radii of curvature of $R = \SI{50.7}{mm}$.
This produces a waist $\omega_0$ and transverse mode spacing $\fTMS$ of:
\begin{equation}
\begin{split}
	& \omega_{0} = \sqrt{\frac{\lambda\sqrt{\epsilon L/2}}{\pi}}  \simeq \SI{49}{\mu m} \\
	& \fTMS = \frac{c}{\pi L}\sqrt{\frac{\epsilon}{R}} \simeq \SI{133}{MHz},
\end{split}
\end{equation}
where $\epsilon = R-L/2 \simeq \SI{1}{mm}$,
 $\lambda = \SI{1064}{nm}$ is the laser wavelength,
 and $c$ is the speed of light~\cite{book:Lasers}.

The nominal frequency difference between the TEM00 and TEM02 or TEM20 modes is \SI{266}{MHz}.
In practice, the horizontal and vertical radii of curvature are slightly different,
 and the resonant frequencies of the TEM02 and TEM20 modes are separated by a few MHz.

%%%%%%%%%%%%%%%%%%%%%%
%\subsection{Control Scheme}\label{sec:cs}

The readout and control scheme is shown in Fig.~\ref{fig:exp}. The laser frequency is locked
to the cavity TEM00 mode, with a 65~kHz bandwidth, using Pound-Drever-Hall reflection locking. 
This servo suppresses laser frequency and cavity length fluctuations that are
common to the three modes. The two frequency shifted beams are then controlled to track
the TEM02 and TEM20 mode resonances so that they probe the sample's coating thermal noise,
which is spatially independent between the three modes. 
In this improved version of the experiment, the higher-order mode probe beams are
controlled using side-of-fringe locking on the cavity transmission. 
To maximize the signal-to-noise ration of these loops, the probe beams are locked at the 
point where the transmission of the TEM02 and TEM20 modes are 70\% of their maximum values. 
Feedback is applied to the two voltage-controlled oscillators (VCO) that determine the frequency
shift of the probe beams, with a control bandwidth of 40~kHz.

With the probe beam frequencies thus slaved to the TEM02 and TEM20 mode frequencies of
the cavity, the spatially independent coating thermal noise of the sample appears in the
frequency difference between the probe beams. This frequency difference is measured by
interfering the two beams, and tracking the fluctuations in the 4~MHz beat signal using
another VCO in a phase-locked loop configuration. The beat signal frequency fluctuations
are converted to an equivalent cavity length change (for the TEM00mode) by multiplying
by the factor $L \lambda /c$. The ASD of this scaled signal, labelled $N_{02/20}$, 
contains the coating thermal noise $N_{\rm CTN}$, as well as other readout noises which
are relatively small in the frequency band of interest.

%%%%%%%%%%
\begin{figure*}[ht]
\includegraphics[width=1.0\textwidth]{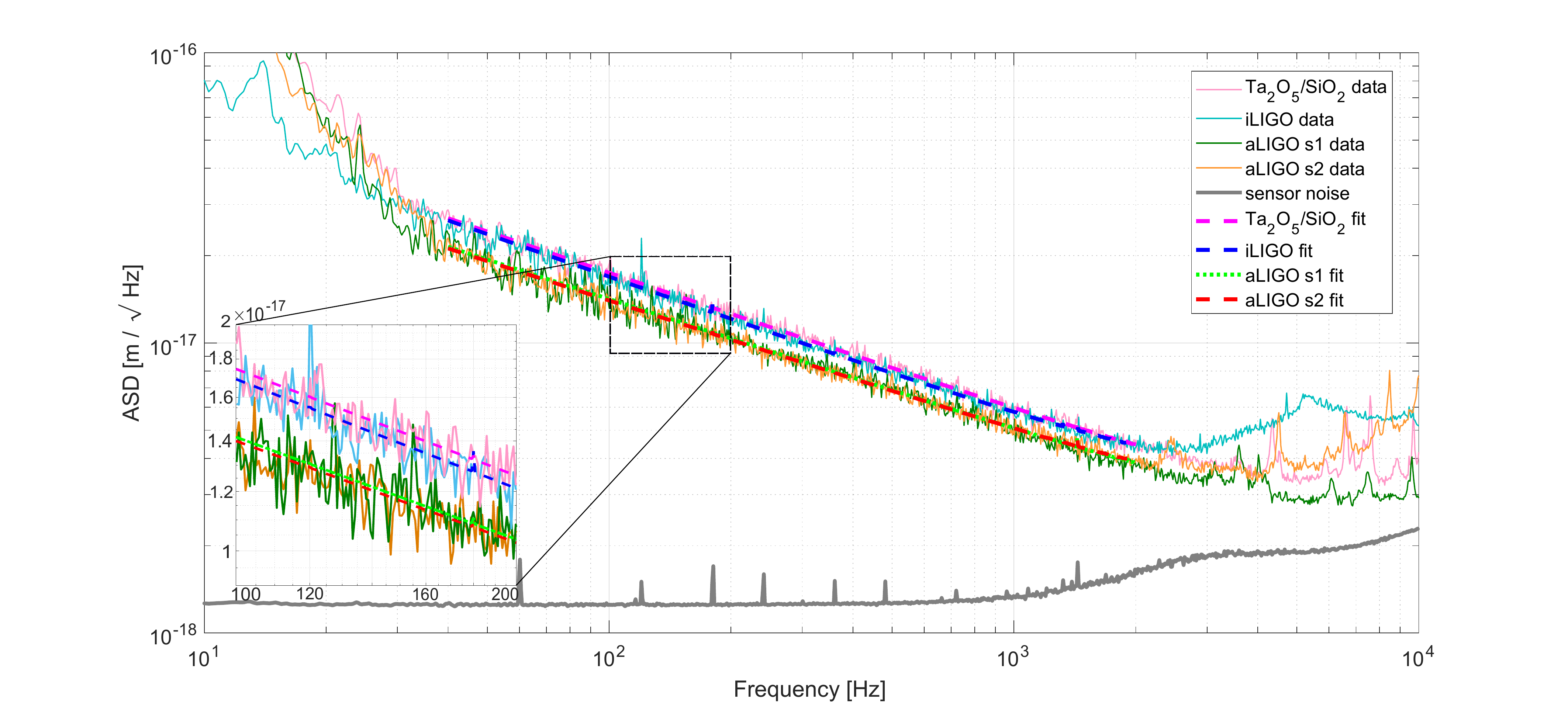}
 \caption{\label{fig:spec} The noise spectrum measured for 3 samples.
 Note that the plotted fit is the sum of CTN and the stationary noise contributions.
 Non-stationary noise below \SI{30}{Hz} from environmental vibrations
  and above \SI{2}{kHz} from down-converted radio frequency (RF) interference,
  limit the extent of the fit.
} 
\end{figure*}
%%%%%%%%%%

These dominant noise sources  are described in the following paragraphs. The VCO used to measure the frequency difference between the higher order modes 
has a noise level of $N_{\rm VCO} \simeq \SI{3}{mHz / \rtHz}$ below 1\,kHz.
This will appear in the readout as an equivalent cavity length noise of:
\begin{equation}
	N_{02/20}^{\rm VCO}(f) = \frac{\lambda L}{c} \, N_{\rm VCO}(f) \simeq \SI{10^{-18}}{\frac{m}{\rtHz}}.
\end{equation}
The VCO noise has some frequency dependence, increasing by about a factor of 2 above $\SI{1}{kHz}$,
as shown in Fig.~\ref{fig:spec}.

The side-of-fringe locking used for the higher-order mode control can be contaminated by fluctuations
in the transmission photocurrents due to both laser intensity noise and shot noise.~The shot noise
associated with the $\SI{400}{\mu W}$ of transmitted power in each higher-order mode corresponds
to a relative intensity noise of $\rm{RIN_s} = 2 \times \SI{10^{-8}}{ Hz^{-1/2}}$. 
This results in a readout noise of:
\begin{equation}
N_{02/20} = 0.7 \frac{\lambda}{\mathcal{F}} \times \rm{RIN_s}  \simeq \SI{10^{-18}}{\frac{m}{\rtHz}} \,
\end{equation}
 which is comparable to the VCO noise described above.

To address laser intensity noise, the power in each probe beam is actively stabilized before
being injected into the cavity. Each probe beam is sampled and detected inside the vacuum
chamber, and intensity servos stabilize the light by controlling
the RF power driving the acousto-optic modulators (see Fig.~\ref{fig:exp}). With a bandwidth
of \SI{50}{kHz}, these servos reduce the probe beam relative intensity noise to below 
$2 \times \SI{10^{-8}}{ Hz^{-1/2}}$ at frequencies below $\SI{10}{kHz}$; higher frequency
residual intensity noise is removed from the transmitted light signals with a simple feed-forward 
circuit.

A lower sensor noise  would require modification of the VCO and the power increase in the cavity by increasing cavity finesse.

%%%%%%%%%%%%%%%%%%%%%%%%%%%%%%%%%%%%%%%%%%%%%%%%%
\section{Extrapolation to TEM00 beams}\label{sec:C}

Our experiment measures the thermal noise sensed by TEM02 and TEM20 modes
in a folded cavity (see Fig.~\ref{fig:cavity}), but we are more typically interested in the
thermal noise for the fundamental mode of a linear cavity. Correction factors are thus required for 
the beam size, mode shape, and folded geometry. These correction factors are described in
detail in \cite{Gras-Yam}; to convert the measured CTN amplitude spectral density, $N_{\rm CTN}$, 
to CTN for a TEM00 beam of size $\omega_L$, this correction is:
\begin{equation}\label{eqn:sctn1}
	N_{\rm CTN}^{00}=0.616 \times \left( \frac{\omega_S}{\omega_L} \right) \, N_{\rm CTN} \,,
\end{equation}  
 where $\omega_S$ is the beam size on the sample mirror (see Table \ref{tab_cav}).

%%%%%%%%%%%%%%%%%%%%%%%%%%%%%%%%%%%%%%
\section{Experimental Results}\label{res}

We measured four coating samples: two witness samples from Advanced LIGO end test mass coatings; 
a witness sample from an initial LIGO end test mass coating; a baseline, standard high-reflectivity coating. 
All four coatings where produced by ion-beam sputtering. The initial LIGO and baseline coatings are 
stacks of quarter-wave Ta$_2$O$_5$-SiO$_2$ doublets. 
For the Advanced LIGO coatings, the Ta$_2$O$_5$ is doped with 25\% TiO$_2$ to reduce mechanical loss~\cite{0264-9381-24-2-008}. 
The layer thicknesses are also altered to further reduce thermal noise: the SiO$_2$ layers are a little thicker and the
Ti-Ta$_2$ layers are a little thinner than a quarter-wavelength.
All sample mirrors have a transmissivity less than \SI{10}{ppm} at \SI{1064}{nm}.   

The baseline coating was deposited at \SI{120}{^\circ}{C}, with a deposition rate of \SI{1.9}{\AA/s} for both materials. 
The sample was then annealed at \SI{450}{^\circ}{C} for 3 hours. The LIGO coating samples were also annealed,
but other coating process parameters for these samples are unknown.

The measured noise, $N_{02/20}$, for all 4 samples are shown in Fig.~\ref{fig:spec}.
In our previous paper we assumed the coating mechanical loss was constant in
frequency, and thus a $1/\sqrt{f}$ coating thermal noise ASD.
With the increased sensitivity of the current experiment, we are able to
measure CTN over a much broader frequency range (\SI{30}{Hz} - \SI{2}{kHz}),
which allows us to measure this slope.
We find that the best fit slope for all samples is near $f^{-0.45}$, which appears to match
 the frequency dependence of the loss angles found in \cite{Amato}.

The fit to the noise spectra for the Advanced LIGO coating samples is:
$$N_{\rm CTN}^{\rm aL}  = {(14.0\pm 0.2)\times10^{-18}}
 \left( {\frac{\SI{100}{Hz}}{f}} \right)^{0.45 \pm 0.02} \frmrtHz. $$

Our fit is limited to the band \SI{30 -- 2000}{Hz}, to avoid the variable environmental 
noise at low frequencies, to remain well above the readout noise floor, and to avoid
small noise peaks at higher frequencies due down-converted radio frequency (RF) interference.

As expected, the other coating samples we measured have higher CTN, since they are
 simple SiO$_2$ and Ta$_2$O$_5$ quarter-wave stacks.
The initial LIGO coating sample has 19\% higher CTN than the the Advanced LIGO coating:
$$N_{\rm CTN}^{\rm iL} = {(16.7\pm 0.1)\times10^{-18}}
 \left( {\frac{\SI{100}{Hz}}{f}} \right)^{0.47 \pm 0.01} \frmrtHz \, $$
while the standard Ta$_2$O$_5$-SiO$_2$ coating has 25\% higher CTN than the
 Advanced LIGO coating:
$$N_{\rm CTN}^{\rm Ta} = {(17.5\pm 0.1)\times10^{-18}}
 \left( {\frac{\SI{100}{Hz}}{f}} \right)^{0.47 \pm 0.03} \frmrtHz. $$
These are consistent with a larger mechanical loss angle for Ta$_2$O$_5$
 without the TiO$_2$ doping.

%%%%%%%%%%%%%%%%%%%%%%

%%%%%%%%%%
\begin{figure}[t!]
\includegraphics[width=0.5\textwidth]{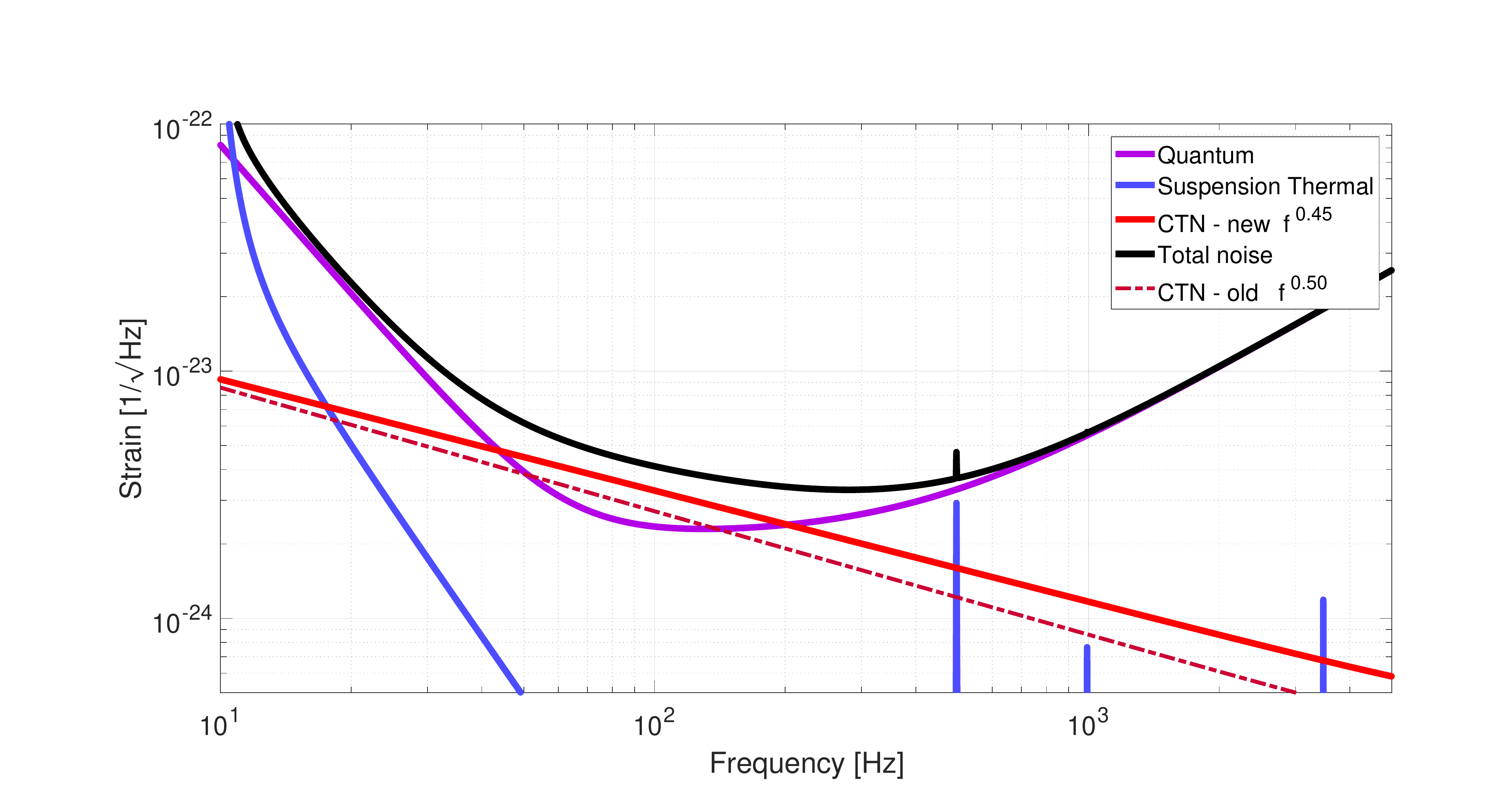}
 \caption{\label{fig:GWINC} The noise budget for aLIGO that 
 incorporates a new measured value of the  loss angle and the slope for coating thermal noise. A previous 
 estimate of coating thermal noise ($\phi_{\rm SiO2} = 5.0\times10^{-4}$, $\phi_{\rm Ti:Ta} = 2.3\times10^{-4}$, 
 slope index = 0.5) is included in the plot and marked as ``CTN-old''. } 
\end{figure}
%%%%%%%%%%%

%\subsection{Variation among  aLIGO samples}
The individual measurements of the two Advanced LIGO coating
samples give the same slope, but slightly different levels of CTN.
At \SI{100}{Hz}, one sample shows $13.9\pm 0.1$ and the other shows
$13.9\pm 0.1$, both in units of $\times10^{-18} \rm{m/Hz^{1/2}}$.
Each sample was measured multiple times at several locations on the coating
and the results where within the statistical error bars.
The CTN difference between the two samples is only 2\%, but it is statistically 
significant (about 3 $\sigma$). 
The origin of this difference is not known, so we extend the uncertainty on
our reported value of $N_{\rm CTN} = {(14.0\pm 0.2)\times10^{-18}} \rm{m/Hz^{1/2}}$ 
to include both measurements.
 
This value differs from our previous estimate $\rm N_{CTN}^{'} = (12.9\pm 0.6)\times10^{-18} \rm{m/Hz^{1/2}}$ \cite{Gras-Yam} 
by less than $2 \sigma$.  The difference may be due in part to small systematic effect resulting from the new experimental set-up,
 or it may simply be due to statistics.
Our previous measurement had an SNR of only 2 at \SI{40}{Hz} (and smaller at other frequencies), 
and the fitting process assumed a white readout noise, so differences at the few percent level 
are not surprising.

%%%%%%%%%%%%%%%%%%%%%%
\subsection{Implications for Advanced LIGO}\label{sec:aligo_ctn}

Extrapolating our measured CTN to the CTN of a \SI{6.2}{cm} beam on an
 Advanced LIGO end test mass using Eqn.~\ref{eqn:sctn1} gives
\begin{equation}
 N_{\rm CTN}^{\rm 00} (\SI{100}{Hz}) = {(7.5\pm 0.1)\times10^{-21}} \frmrtHz .
\end{equation}
This is slightly higher than our previously reported value, and higher than the value
 used in Advanced LIGO design documents
 ($5.8 \times10^{-21} \mrtHz$ at \SI{100}{Hz} \cite{0264-9381-32-7-074001}).
Using the CTN value and slope measured here, we find an overall decrease 
in the expected Advanced LIGO binary neutron star range of 7\% (from \SI{186}{Mpc} to \SI{171}{Mpc}~\cite{Chen:2017aa})
compared to \cite{0264-9381-32-7-074001, den_nb}, see Fig. \ref{fig:GWINC}.

%%%%%%%%%%%%%%%%%%%%%%  
\subsection{Loss angle of \ttan:\tant}

To estimate the loss angle for the titania-tantala alloy used as the high refractive index material
in the Advanced LIGO coatings, we use the equations given in \cite{PhysRevD.87.082001}
and assume a  loss angle for silicon-dioxide (the low index material) of 
$\phi_{\rm Si02} = 5\times10^{-5}$ \cite{PhysRevD.91.022005}.
We further assume that the loss angles associated with shear and bulk deformation are equal
in both coating materials.
We have moved away from the simplified CTN equations from \cite{PhysRevD.91.042002}
used in our previous publication because that calculation neglects field penetration into the coating
 and thus underestimates the loss angle of the high index material by 4\%.
The current experiment's precision is sufficient to make this a non-negligible effect.

Our estimate for the loss angle of the high-refractive-index material
 in the Advanced LIGO coatings is 
\begin{equation}
\phi_{\rm Ti:Ta}={(3.6\pm 0.1})\times 10^{-4} \left( {\frac{f}{\SI{100}{Hz}}} \right)^{0.1 \pm 0.04}.
\end{equation}
This number is slightly lower than the value previously reported in \cite{PhysRevD.91.022005},
 but higher than the values reported in \cite{0264-9381-27-8-084030, Gras-Yam}.

Using the same procedure, we estimate the loss angle of tantala in the Ta$_2$O$_5$-SiO$_2$ coatings.
We obtain the same value for both coatings, 
\begin{equation}
\phi_{\rm Ta}={(5.3\pm 0.1})\times 10^{-4} \left( {\frac{f}{\SI{100}{Hz}}} \right)^{0.06 \pm 0.02},
\end{equation}
 which is higher than reported in \cite{0264-9381-20-13-334,0264-9381-23-15-014}.

\vspace{20pt}
%%%%%%%%%%%%%%%%%%%%%%%%%%%%%%%%%%%%%%
\section{Conclusions}

Precision measurements of coating thermal noise are critical to 
 both high-precision laboratory-scale R\&D, and large scale efforts such as gravitational-wave detectors.
Our finding that the CTN spectrum deviates from the assumed slope will allow for more
 reliable computations of CTN from measurements of the mechanical properties,
 and more accurate extrapolations of direct CTN measurements to other frequency bands.

For Advanced LIGO in particular, the measurements presented allow us to update
 our understanding of the sensitivity achievable by current detectors.
The CTN estimated for Advanced LIGO from our measurements is higher than 
that originally computed for Advanced LIGO, and it results in a 7\% reduction in the detectors' 
expected range.
Similar impacts are expected for other gravitational-wave detectors,
and both the amplitude and slope of CTN measured here will need to be incorporated
into future detector designs.

%%%%%%%%%%%%%%%%%%
\begin{acknowledgments}
The authors would like to acknowledge the unfailing support and recognition
 of the LIGO Scientific Collaboration's optics working group
 without which this work would not have been possible. ​The authors also acknowledge the support of
 the National Science Foundation under Grant 6936650.
We are also very grateful for the computing support provided by The MathWorks, Inc.

LIGO was constructed by the California Institute of Technology and
 Massachusetts Institute of Technology with funding from the National Science Foundation,
 and operates under cooperative agreement PHY-0757058.
Advanced LIGO was built under award PHY-0823459.
This paper carries LIGO Document Number LIGO-P1700448.
\end{acknowledgments}

\vspace{20pt}
\def\LSC{{The LIGO Scientific Collaboration}}
\bibliography{CTN2}% Produces the bibliography via BibTeX.

\end{document}